\documentclass{aims}

\usepackage{graphicx}
\usepackage{txfonts}
\def\typeofarticle{Research article}
\def\currentvolume{21}
\def\currentissue{x}
\def\currentyear{2024}

\def\ppages{xxx--xxx}
\def\DOI{}
\def\Received{03 July 2024}
\def\Revised{17 August 2024}
\def\Accepted{26 August 2024}
\def\Published{xxxx}

\numberwithin{equation}{section}

\begin{document}

\title{Inference of a Susceptible–Infectious (SI) stochastic model }

\author{%
 Giuseppina Albano\affil{1,}\corrauth, \
Virginia Giorno\affil{2} 
and
Francisco Torres-Ruiz\affil{3,4}
}

\shortauthors{the author(s)}

\address{%
  \addr{\affilnum{1}}{Dipartimento di Studi Politici e Sociali, Università degli Studi di Salerno, Via Giovanni Paolo II, 84084 Fisciano (SA), Italy}
  \addr{\affilnum{2}}{Dipartimento di Informatica,  Università degli Studi di Salerno, Via Giovanni Paolo II, 84084 Fisciano (SA), Italy}
  \addr{\affilnum{3}}{Departamento de Estad\'istica e I.O., Universidad de Granada, Avenida de Fuente Nueva s/n, 18071, Granada, Spain}
\addr{\affilnum{4}}{Instituto de Matem\' aticas de la Universidad de Granada (IMAG), Calle Ventanilla 11, 18001, Granada, Spain}}

\corraddr{Email: pialbano@unisa.it; Tel: +39-089-962-645.
}

\begin{abstract}
We consider a time-inhomogeneous diffusion process able to describe the dynamics of infected people in a susceptible-infectious epidemic model in which the transmission intensity function is time-dependent. Such a model is well suited to describe some classes of micro-parasitic infections in which individuals never acquire lasting immunity and over the course of the epidemic everyone eventually becomes infected. The stochastic process related to the deterministic model is transformable into a non homogeneous Wiener process so the probability distribution can be obtained. Here we focus on the inference for such process, by providing an estimation procedure for the involved parameters. We point out that the time dependence in the infinitesimal moments of the diffusion process makes classical inference methods inapplicable. The proposed procedure is based on Generalized Method of Moments in order  to find suitable estimate for the  infinitesimal drift and  variance of the transformed  process. Several simulation studies are conduced to test the procedure, these include the time homogeneous case, for which a comparison with the results obtained by applying the MLE is made, and cases in which the intensity function are time dependent with particular attention to periodic cases. Finally, we apply the estimation procedure to a real dataset.

\end{abstract}


\maketitle

\section{Introduction}
The onset of large epidemics in recent decades, such as the SARS epidemic, the avian influenza, Ebola  and  the Covid-19 pandemic, have meant that a lot of  mathematical models to study various infectious diseases have been developed.
(see, for example, \cite{Allen_2010}--\cite{SIModel_2022}).
The main aim of such studies is to forecast the dynamics of the  disease,  by using also  suitable inference techniques,  in order to adopt  appropriate containment policies (see \cite{Tizzoni_2012}--\cite{Albano_2021}).
\par
Generally, in epidemic models the population is divided into compartments that constitute a partition of it, that is, each compartment constitutes a subset of the population disjoint from the others and the union of all the compartments returns the whole population. Among the models that contemplate this compartmentalized philosophy, the SIR type and its derivatives stand out.

The study of disease propagation through these models, both in their deterministic and stochastic versions, has experienced a great boom in recent decades, giving rise to an extensive literature. The range of models considered is very wide, taking into account different points of view. For example, in \cite{Zhu21} and \cite{Seb22}, Kalman filtering techniques are applied to estimate the states of a discrete nonlinear compartmental model. In stochastic environment, Markovian models occupy a prominent place. Within them, if we consider models with discrete states, continuous-time Markov chains have been widely used (see, for instance, \cite{Art14}--\cite{Artalejo2010}). As for continuous-time Markovian models with continuous state space, diffusion processes arise naturally by introducing random environment (through a multidimensional Wiener process) into the systems of ordinary differential equations governing deterministic models.

From the classical SIR model, a number of increasingly complex variants have emerged. Among the lines on which the evolution of the models has been based, we can highlight the following:
\begin{itemize}
  \item The partitioning of the total population into an increasing number of compartments, each of which obeys a particular situation of the individuals. Thus, in addition to the usual susceptible (S), infected (I) and recovered (R) individuals, there are others such as those with passive immunity (M), the exposed and uninfected (E), deceased (D), among others. This has led to increasingly complex and difficult to treat compartment models (see for instance \cite{Papa2023,Papa2024}).

  \item The inclusion of the mechanisms of contagion of the disease, the so-called incidence function. The most usual function of this type is the one that relates susceptible individuals to infected individuals according to the law of mass action, but considering a constant contagion rate and/or rational functions of both groups of individuals to regulate the contagion scheme (\cite{Raj19,Li21}).

  \item The inclusion of terms describing the effect of vaccination of individuals (including the possibility of restrictions in the vaccination process), as well as the possible effects of cross-infection (see, for instance \cite{Xue22}--\cite{Cha22})

\item The emergence of new diseases. The effect of COVID19 on the increase in the literature on this type of model cannot be denied. Indeed, the analysis of the effects of the pandemic has given rise to numerous publications focusing on two lines of action: on the one hand, the study of the evolution of the disease in specific locations using existing models (see \cite{Bod22}) and, on the other hand, the appearance of new models (see, for example, \cite{Leitao}).
\end{itemize}

However, it should be noted that a very high number of these publications focus their interest on the description of the model and its analysis from a theoretical point of view (existence, uniqueness and non-explosion of the solution, stability and equilibrium problems, extinction of the disease). However, especially in stochastic models, the aspects derived from the estimation of the models are not yet well developed, resorting to simulations and/or Monte Carlo type methods to illustrate the validity of the proposed models. The main reason for this derives from the difficulty to know
the probability distribution governing the dynamic evolution of the model (more complex the more complicated the model).
Some approximation has been made on simpler models (such as the SIS) by means of approximations derived from Euler-Maruyama type discretizations (see, for instance \cite{Pan14} ), although this requires certain conditions on the
model components (e.g. the incidence function). The situation can become even more complex if the rates involved in the model are not considered constant but time-dependent.

This paper therefore falls along the line of inference in epidemic models. The aim is the inference of a Susceptible-Infected (SI) type model, which is essentially a \lq\lq simplified\rq\rq\  version of a SIR type model. Precisely, in the SI model it is assumed that an individual can be in one of only two states, either susceptible ($S$) or infectious ($I$). Although this model is quite simple, it is adept at capturing several types of diseases in which individuals remain infected for life (e.g. brucellosis in domestic and wild populations, fox rabies). Even the most famous AIDS has been modeled in the literature using an SI-type model (see, for example, \cite{Pugliese90}).

For $t \geq t_0$, we denote by $S(t)$ the number  of susceptible individuals,  by $I(t)$ the number of individuals infected and by $K$ the total  population size, where $K = S(t) + I(t)$ is constant.  In similar  models, infected individuals are lifetime infectious. We point out that here the size of the population is assumed to be constant, i.e. the birth and death rates of both populations S and I are assumed to be negligible. This is a very strong assumption, but reasonable if one observes the phenomenon for a limited time.
\par
A model with two states is described via ordinary differential equations in the deterministic dynamics. More general models can be build making use of stochastic approaches based on the birth and death processes or diffusion processes (see, for instance, \cite{Allen_2010}, \cite{Giorno_2023a,Giorno_2023b} and references therein); they are  more realistic but more complicated to analyze. Such models try to forecast the spread of the disease, in terms of the total number of infected people, the duration of an epidemics. Moreover, they  allow to estimate suitable epidemiological parameters, as the transmission rate of the disease measured via the   basic reproduction number, i.e. the expected number of infected cases directly generated by one case in a population where all individuals are susceptible or infected.
\par
Let $I(t)$ and $S(t)$ be the sizes of the infected  and the susceptible populations, respectively. We consider the  deterministic susceptible-infected (SI) model describes by the following
\begin{eqnarray*}
&&{dS(t)\over dt}=-{\lambda(t)\over K}\,S(t)\,I(t),\\
&&{dI(t)\over dt}={\lambda(t)\over K}\,S(t)\,I(t),
\end{eqnarray*}
where the transmission intensity function $\lambda(t)$ is a positive, bounded and continuous function of $t$. It follows that the population dynamics of the infected $I(t)$ can be described by the Pearl-Verhust logistic growth differential equation:
\begin{equation}
{d I(t)\over dt}={\lambda(t)\over K}[K-I(t)]\,I(t),\qquad t>t_0,
\label{differential_equation}
\end{equation}
where the transmission intensity function $\lambda(t)$ is a positive, bounded and continuous function of $t$. The solution of (\ref{differential_equation}) is
\begin{equation}
I(t)={K\,I(t_0)\over I(t_0)+[K-I(t_0)]\,e^{-\Lambda(t|t_0)}},\qquad t\geq t_0,
\label{solution}
\end{equation}
with
\begin{equation}
\Lambda(t|t_0)=\int_{t_0}^t\lambda(\theta)\;d\theta.
\label{int_lambda}
\end{equation}
We note that $\lim_{t\to\infty} I(t)=K$, so the whole population is destined to become infected, hence the parameter $K$ identifies the carrying capacity of the infected population.  We point out that the susceptible-infectious epidemic model  is an extreme case of the more general models including recovered population and can be obtained from them by assuming that the time required to reach an immunity situation is infinitely long (see, for instance, \cite{Ram21}). The state $K$ is
not reachable in finite time, so that it is  interesting to consider  the time $T^*_m$ required to reach a fixed threshold $m$. When the process is time-homogeneous , i.e. the function $\lambda(t)$ is constant, one can determine the time $T^*_m$  such that $I(T^*_m)=m$. In particular, from \eqref{solution} one obtains
$$
	T^*_m=t_0+\frac{1}{\lambda}\,\ln\frac{K-I(t_0)}{(K-m)\,I(t_0)},
$$
being $I(t_0)$ the initial size of the infected population and $m\in(0,K)$. \par
To generate a stochastic diffusion process for $I(t)$ various approaches can be used, generally, they  introduce  stochastic elements into the equation \eqref{differential_equation} or into solution \eqref{solution}. These two approaches lead to different processes described by stochastic differential equations of Itô or Stratonovich type, respectively (see \cite{Arnold_1974}). In this work, the first of the two approaches has been chosen.
\par
In \cite{Giorno_2023b} a time-inhomogeneous diffusion process to model the size of the infected population in a stochastic environment is provided by starting from \eqref{solution}. For such a process, the authors study  the  probability distribution and derive closed form results for the  first-passage time problem through a constant boundary  to obtain the stochastic counterpart of the parameter $T^*_m$. However, such important issues as that of model inference were not addressed in this study.
\par
The aim of the present  paper is to provide  the inference on  the stochastic diffusion process built from  \eqref{solution}. Indeed such issue becomes fundamental to study the dynamics of infectious diseases and  to measure their aggressiveness so  to make predictions about future infections.
\par
The paper is organized as follows. Section~2 provides the stochastic model and its probability distribution, i.e. the transition probability density probability (pdf) function and the related moments. Section 3 contains the inference procedure based on the Generalized Method of Moments (GMM) for fitting the parameters and the unknown functions of the model. In Section~4 some simulation experiments are performed to validate the proposed procedure.
An application to real data is also considered in Section~5. Some conclusions close the paper.

%
\section{The model}
Under the assumption of random environment, let $\{X(t),t\geq t_0\}$  be the stochastic process describing the size of the infected population at time $t$, and
we interpret $\Lambda(t|t_0)$ as the mean of a time-inhomogeneous Wiener process $\{Z(t),t\geq t_0\}$, described by the stochastic equation
\begin{equation}
Z(t)=\Lambda(t|t_0)+W\bigl[V(t|t_0)\bigr],\qquad t\geq t_0,
\label{Zt}
\end{equation}
where $W(t)$ is the standard Wiener process and
\begin{equation}
V(t|t_0)=\int_{t_0}^t\sigma^2(\theta)\;d\theta, \qquad t\geq t_0,
\label{Vt}
\end{equation}
with  $\sigma(t)$  positive, bounded and continuous function of  $t$ describing  the breadth of the random oscillations.
Hence, Eq. \eqref{solution} is generalized by the following stochastic equation:
\begin{equation}
X(t)={K\,X(t_0)\over X(t_0)+[K-X(t_0)]\,e^{-Z(t)}},\qquad t\geq t_0.
\label{stochasticsolution}
\end{equation}
As proved in \cite{Giorno_2023b}, $\{X(t),t\geq t_0\}$ is a time-inhomogeneous diffusion process defined in $(0,K)$ described by the following stochastic differential equation:
\begin{equation}
dX(t)= A_1(x,t)\,dt+ \sqrt{A_2(x,t)}\, dW(t),\qquad X(t_0)=x_0, \label{stoc_equation}
\end{equation}
where
\begin{equation}
A_1(x,t)={\lambda(t)\over K}(K-x)x
+{1\over 4}{\partial A_2(x,t)\over \partial x}
={(K-x)x\over K} \Bigl[\lambda(t)+\frac{1}{2}\,\sigma^2(t)\Bigr],\qquad A_2(x,t)=\sigma^2(t)\,{(K-x)^2x^2\over K^2},
\label{inf_moments}
\end{equation}
are the infinitesimal drift and the infinitesimal variance of $X(t)$, respectively, and $x_0\in (0,K)$ is the initial size of the infected population.
\par
As shown in \cite{Giorno_2023b}, the process $X(t)$ can  be transformed into a time-inhomogeneous Wiener process $Y(t)$
with drift and infinitesimal variance
\begin{equation}
B_1(t)=\lambda(t), \qquad B_2(t)=\sigma^2(t),
\label{MomentiInfinitesimaliWiener}
\end{equation} by using the transformation:
 \begin{equation}
y=\int_{x_0}^x {K\,dz\over z(K-z) }=\ln\Biggl[{x(K-x_0)\over x_0(K-x)}\Biggr],\quad y_0=0.
 \label{transformations}
 \end{equation}
We point out that $Y(t)$ is a Gauss Markov process with  mean
\begin{equation}
{\mathbb E}[Y(t)]=\Lambda(t|t_0)
\label{MediaY}
\end{equation}
and covariance function
\begin{equation}
{\rm cov}\big[Y(s), Y(t)\big]=V(s|t_0),\quad t_0\leq s\leq t.
\label{covarianceY}
\end{equation}
Further, since the transition pdf $f_Y(y,t|y_0,t_0)$ of  $Y(t)$ is normal with mean $y_0+\Lambda(t|t_0)$ and variance $V(t|t_0)$, we can obtain the transition pdf of the process $X(t)$:
\begin{equation}
\hspace*{-0.3cm}
f_X(x,t|x_0,t_0)={K\over x(K-x)}\,{1\over\sqrt{2\pi\,V(t|t_0)}}\exp\Biggl\{ - {\bigl[\ln\bigl({x(K-x_0)\over x_0(K-x)}\bigr)-\Lambda(t|t_0)\bigr]^2\over 2\,V(t|t_0)}\Biggr\}.
\label{pdf_X}
\end{equation}
 Moreover, the transition distribution function of $X(t)$ is
\begin{equation}
F_X(x,t|x_0,t_0)={1\over 2}\, \biggl\{ 1+{\rm Erf} \biggl[ {\ln\bigl({x(K-x_0)\over x_0(K-x)}\bigr)-\Lambda(t|t_0)\over \sqrt{2\,V(t|t_0) }}\biggr]\biggr\},\qquad x,x_0\in(0,K).
\label{distribution_X}
\end{equation}
where ${\rm Erf}(z)=\frac{2}{\sqrt \pi}\int_0^z e^{-u^2}du$ is the error function. \par
The conditional median  $\mu[X(t) |X(t_0)=x_0]$  of the process $X(t)$ can be obtained  from (\ref{distribution_X}) by imposing  that $F_X(x,t|x_0,t_0)=1/2$ for $t\geq t_0$, so that one has:
\begin{equation}
\mu[X(t) |X(t_0)=x_0]={K\,x_0\over x_0+(K-x_0)\,e^{-\Lambda(t|t_0)}},\qquad t\geq t_0.
\label{cond_median}
\end{equation}
Moreover, for  $m=1,2,\ldots$, the $m$-th conditional moment of $X(t)$ is:
\begin{equation}
{\rm E}[X^m(t)| X(t_0)=x_0]={K^m\over\sqrt{\pi}}\,\int_{-\infty}^{+\infty} \Biggl[1+{K-x_0\over x_0}\,\exp\Bigl\{-z\sqrt{2V(t|t_0)}-\Lambda(t|t_0)\Bigr\}\Biggl]^{-m}\,e^{-z^2}\;dz.
\label{moments_SI_diff}
\end{equation}
Such analytical results will be used in the following section to make inference for the process $X(t)$, to provide a fitting procedure based on GMM.
 %
\section{Inference}
The transformation \eqref{transformations} is the basis of our estimation procedure for the functions $\lambda(t)$ and $\sigma^2(t)$. The idea is to estimate such functions by making inference on the transformed Wiener process $Y(t)$ in order to fit $\lambda(t)$ and $\sigma^2(t)$. In particular, from \eqref{MediaY} and \eqref{covarianceY} we obtain:
\begin{eqnarray}
&&\lambda(t)=\frac{d\, {\mathbb E}[Y(t)]}{dt},\label{MY}\\
&&\sigma^2(s)=\frac{d\, {\rm cov}[Y(s),Y(t)]}{ds}.\label{CovY}
\end{eqnarray}
In the following we assume that the carrying capacity $K$ is known since it represents the total size of the population, whereas the functions to be estimated are $\lambda(t)$ and $\sigma^2(t)$, for  $t$ belonging to  the observation interval $[t_0,T]$.  \\
We consider a discrete sampling of the process \eqref{stoc_equation} based on $d$ sample paths observed at the times $t_{j}$, with $j=1,\ldots,n$. For $i=1,\ldots,d$, let $x_{ij}$ be the observed values of the $i$-th path at times $t_{j}$,  the values $x_{i1}$ represent  the initial point of the sample paths. \\
The inference procedure is illustrated in the following:
\begin{itemize}
\item From the observed data $x_{ij}$ for $ i=1,\ldots,d$ and  $j=1,\ldots,n$ of the process $X(t)$, obtain the points $y_{ij}$ as following:
\begin{equation}
y_{ij}=\ln\Biggl[{x_{ij}(K-x_{i1})\over x_{i1}(K-x_{ij})}\Biggr]
\label{tranformed_points}
\end{equation}
Such points can be considered as observations of the Wiener process $Y(t)$ given in \eqref{MomentiInfinitesimaliWiener}.
\item From the data  $y_{ij}$ for $i=1,\ldots,d$,  obtain the sample mean $\mu_j$
\begin{equation}
\mu_j=\frac{1}{d}\sum_{i=1}^d y_{ij} \qquad (j=1,\ldots,n)
\end{equation}
and the sample covariance $\nu_j$ between two subsequent observations $Y(t_{j-1})$ and $Y(t_{j})$:
\begin{equation}
\nu_j=\frac{1}{d-1}\sum_{i=1}^d (y_{ij}-\mu_j) (y_{i\,j-1}-\mu_{j-1}),\qquad j=2,\ldots,n.
\end{equation}
\item Interpolate the values $\mu_j$ for $j=1,\ldots,n$ and $\nu_j$ for $j=2,\ldots,n$. Let $\widehat M(t)$ and $\widehat \Sigma(t)$ be the functions obtained.
\item Evaluate the derivatives of $\widehat M(t)$ and $\widehat\Sigma(t)$.
\item From Eq. \eqref{MY} and \eqref{CovY}  obtain the estimate of $\lambda(t)$ and $\sigma^2(t)$ as follows:
\begin{equation}
\widehat\lambda(t)=\frac{d\, {\widehat M}(t)}{dt}\qquad \widehat\sigma^2(t)=\frac{d\, \widehat \Sigma(t)}{dt}.\label{stimatori}
\end{equation}
\end{itemize}
We point out that the  procedure just illustrated  combines  a GMM estimation with  the interpolation of the sample mean and covariance points, so
the consistence of the  estimators \eqref{stimatori} of $\lambda(t)$ and $\sigma^2(t)$ derives from the consistence of the GMM estimator and from  the uniform convergence of the interpolation method, for example in our analysis we can consider cubic spline interpolation. \\
Finally, we note that in real applications, it could happen that the observed paths do not reach the carrying capacity since the phenomenon is observed before $K$ is achieved.
In these cases, we argue that an a priori rough estimate of the parameter $K$  can be  obtained by starting from the maximum observed point. Such estimate  can be used to obtain the points $y_{ij}$ for $i=1,\ldots,d$ and $j=1,\ldots,n$ from which, by using the previous procedure, we fit the functions $\lambda(t)$ and $\sigma^2(t)$. An improvement of the estimate of $K$ could be then obtained by looking at the conditional median function \eqref{cond_median} and by implementing a step by step procedure  similar to that proposed in the \cite{CSDA2020}. This topic will be the subject of future investigations.

\section{Some simulation experiments}
In order to evaluate the suggested procedure, in the following  we consider several simulation experiments in which $d$ sample-paths of $X(t)$ are simulated, each including $n$ equally
spaced observations in $[t_0 , T ]$ with $t_0=0$, $T = 50$, $t_i - t_{i-1} = \Delta=0.01$  $(i = 1,\ldots ,n)$ and $x_0 = 20$. The estimation of the unknown functions is replicated $N = 500$ times.
In all the cases we choose the carrying capacity $K=200$. In Section 4.1 we consider the case in which both the functions $\lambda(t)$ and $\sigma^2(t)$ are time independent, whereas in the
Section 4.2 one or both of them are continuous functions of the time.
\subsection{The time homogenous case}
\begin{figure}[H]
\centering
\includegraphics[scale=0.37]{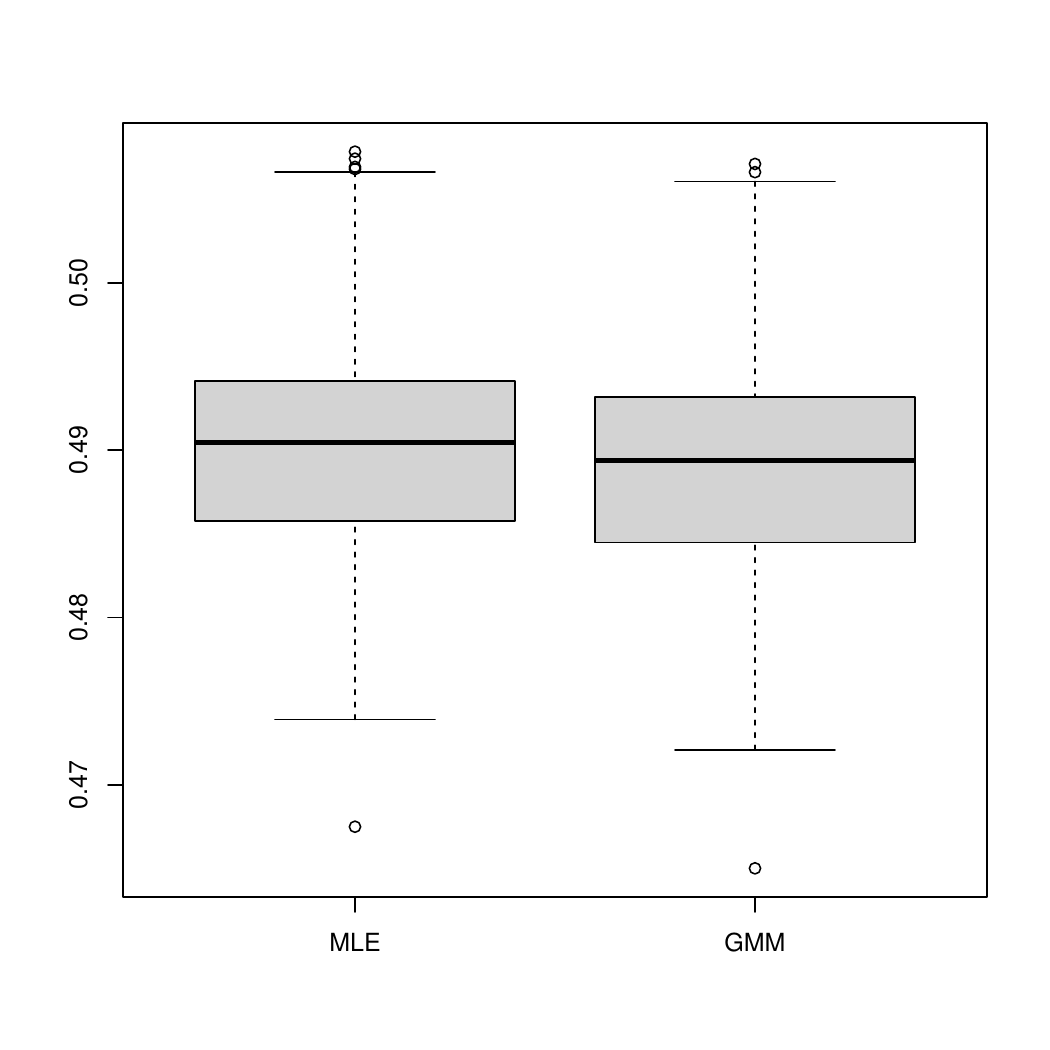}\includegraphics[scale=0.37]{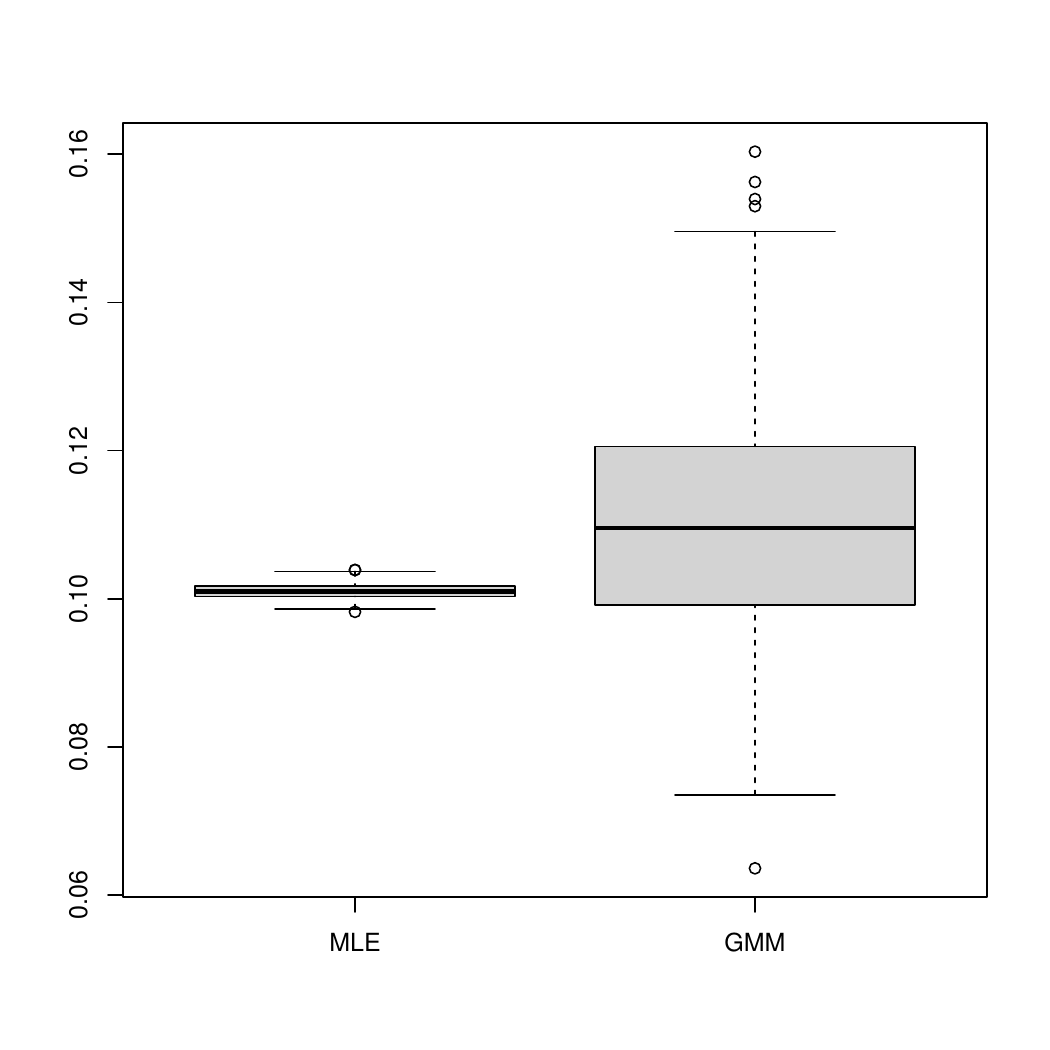}\\
\includegraphics[scale=0.37]{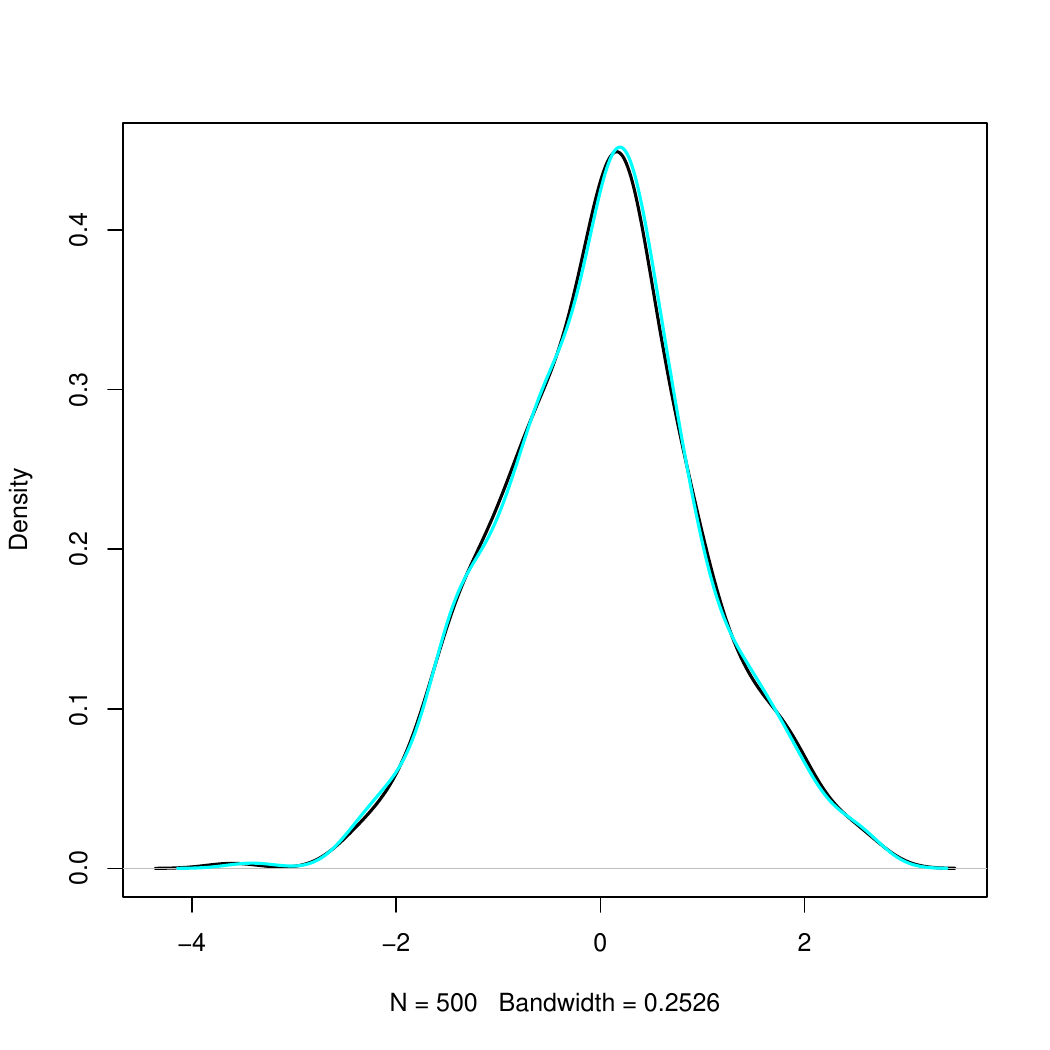}\includegraphics[scale=0.37]{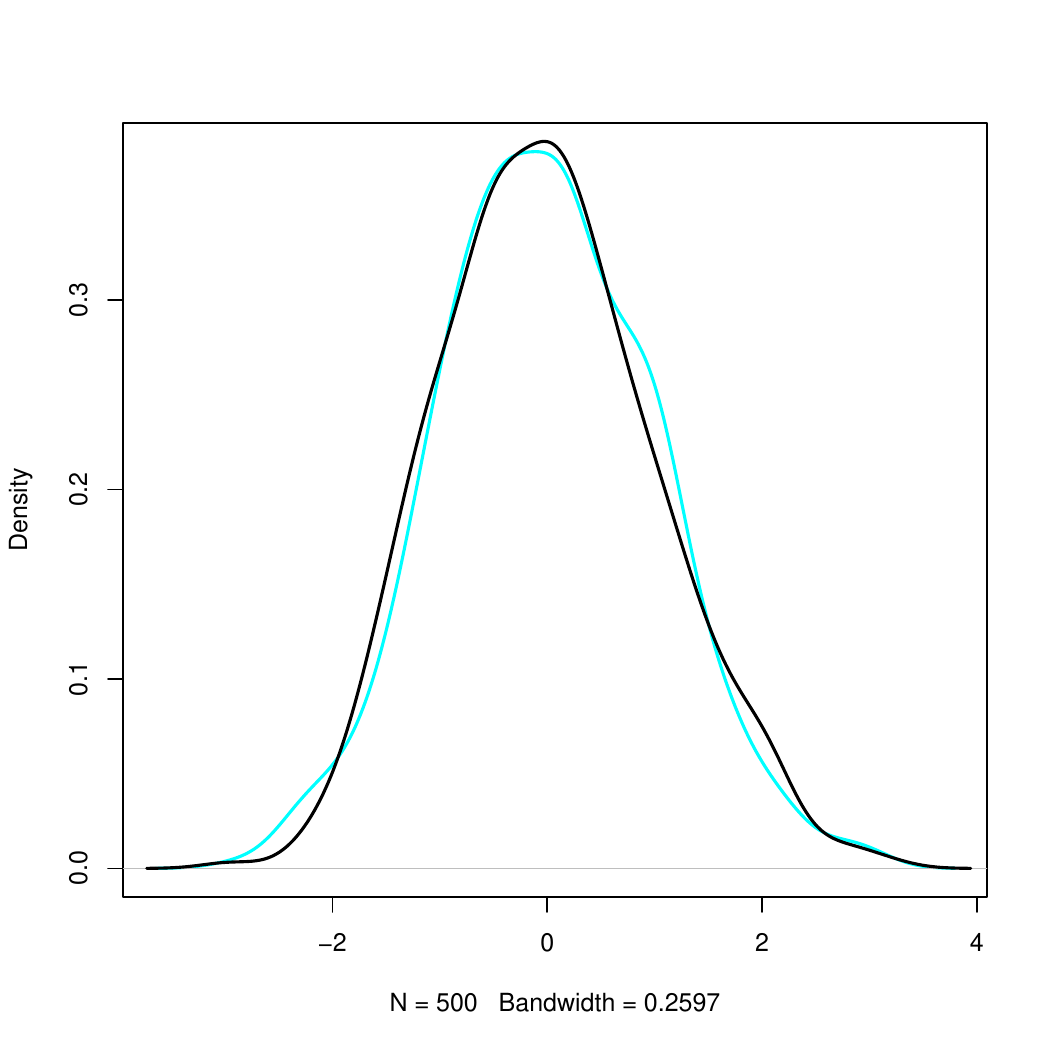}
\caption{On the top: Box plot of MLE and GMM estimates of the parameters $\lambda$ (on the left) and of $\sigma^2$ (on the right) based on 500 replicates. On the bottom: Gaussian kernel density estimates of  $\lambda$ (on the left) and of $\sigma^2$ (on the right). The smoothing bandwidth is also indicated. Black curve is the GMM density, while red curve is the MLE density.}
\label{Fig1}
\end{figure}

Suppose that $\lambda(t)$ and $\sigma^2(t)$ in \eqref{inf_moments} are constant functions. In particular, in this simulation study, we consider
$$
\lambda(t)=\lambda=0.4,\qquad \sigma^2(t)=\sigma^2=0.1.
$$
In this case, the process $X(t)$ in \eqref{inf_moments} is time homogeneous, so the inference for the parameters $\lambda$ and $\sigma^2$ can be  made by means of the classical Maximum Likelihood Estimation (MLE). Indeed, it is easy to obtain the MLE for $\lambda$ and $\sigma^2$ by looking at the Log-Likelihood function of the process $Y(t)$ in \eqref{MomentiInfinitesimaliWiener}:
\begin{equation}
\log L_Y(\lambda,\sigma^2)=-\frac{d(n-1)}{2}\log(2\pi\Delta)-\frac{d(n-1)}{2}\log \sigma^2
-\frac{1}{2\sigma^2\Delta}\sum_{i=1}^d \sum_{j=2}^n (y_{ij}-y_{i j-1}-\lambda\Delta)^2
\end{equation}
In particular, we obtain the following estimator:
\begin{equation}
\widehat\lambda_{MLE}=\frac{1}{d(n-1)\Delta}\sum_{i=1}^d \sum_{j=2}^n (y_{ij}-y_{i j-1}), \qquad \widehat\sigma^2_{MLE}=\frac{1}{d(n-1)\Delta}\sum_{i=1}^d \sum_{j=2}^n (y_{ij}-y_{i j-1}-\widehat\lambda_{MLE})^2.
\end{equation}
In Figures~\ref{Fig1} we compare the results obtained via  the MLE method with those obtained via our procedure in the following denoted by GMM.
In particular, in Figure~\ref{Fig1} on the top the box-plots of the 500 estimates of $\lambda$ (on the left) and of $\sigma^2$ (on the right) are shown. We can see that the estimates of $\lambda$ by means of MLE and GMM have quite the same distribution ranging from $0.46$ to $0.51$. Differently, the estimates of $\sigma^2$ via MLE present a very low variability compared to those of the GMM, resulting that MLE is to be preferred in this case. However, we point out the MLE assumes that the parameters $\lambda$ and $\sigma^2$ are constant and so it looks at the Log-Likelihood as a function of such parameters; instead, our procedure can be applied in the general case in which we do not have such information and the parameters generally depend on time. On the bottom of Figure \ref{Fig1} the kernel density  of the MLE and GMM standardized  estimates of $\lambda$ (on the left) and $\sigma^2$ (on the right) are shown. The red curves represent the density of MLE, whereas the black curves refer to  GMM. We can observe that the sampling distributions of both the  estimators and both the methods are quite symmetric and superimposable.
Finally, let
$$
MRE(\widehat \lambda)=\frac{1}{500}\sum_{i=1}^{500}\frac{|\widehat\lambda_i-\lambda|}{\lambda}, \qquad MRE(\widehat \sigma^2)=\frac{1}{500}\sum_{i=1}^{500}\frac{|\widehat\sigma^2_i-\sigma^2|}{\sigma^2}
$$
be the mean relative errors  of the estimators $\widehat\lambda$ and $\widehat\sigma^2$. In Table~\ref{Tabella1} the MRE's are shown  for different choices of the parameters $\lambda$  and $\sigma^2$, i.e. $\lambda=0.4,0.6,0.8,1.2$ and $\sigma^2=0.05, 0.1$, and for the two considered methods.
The MRE's obtained via the MLE and via  the GMM are generally comparable, especially for $\lambda$,  even though the MLE method provides lower errors as  expected by seeing the box plots in the Figure~\ref{Fig1}. This one is due to the strong assumption of the constancy of the parameters. Moreover, the MRE's increase as $\sigma^2$ increases for both the estimators and for both the methods.

\begin{table}[]
\centering
\caption{Mean relative errors for the MLE and GMM estimators for several choices of the parameters $\lambda$ and $\sigma^2$.}\label{Tabella1}
\begin{tabular}{cccccc}
\hline
  $\lambda$& $\sigma^2$ & MRE$(\widehat\lambda_{MLE})$ &MRE$(\widehat\lambda_{GMM})$ & MRE$(\widehat\sigma^2_{MLE})$ & MRE$(\widehat\sigma^2_{GMM})$\\
 \hline
 0.4& 0.05 & 0.113 & 0.110 & 0.010 & 0.123\\
0.6& 0.05 & 0.078 & 0.075 & 0.014 & 0.115\\
0.8& 0.05 &0.116& 0.119 & 0.213 & 0.175\\
1.0 & 0.05 & 0.282& 0.284 & 0.474& 0.299\\
1.2 & 0.05 &0.396 & 0.398 & 0.712 & 0.389  \\
0.4& 0.1  &  0.225 & 0.223 & 0.011& 0.149\\
0.6& 0.1  & 0.131& 0.128 & 0.023& 0.114\\
0.8& 0.1  &  0.103 & 0.106 & 0.131 & 0.190\\
1.0& 0.1  &  0.271 & 0.273 & 0.225 & 0.315\\
1.2& 0.1 &0.387& 0.389 & 0.303 & 0.406 \\
\hline
\end{tabular}

\end{table}

\subsection{The time inhomogenous case}
In this Section we consider the general case in which one or both the functions $\lambda(t)$ and $\sigma^2(t)$ are time dependent.
In the following we analyze three different cases:
\hspace{0.5cm}
\begin{itemize}
\item[{\bf a.}] $\lambda(t)=0.4+\sin t, \qquad \sigma^2(t)=\sigma^2=0.1$;
\item[{\bf b.}]  $\lambda(t)=0.4+\sin t, \qquad \sigma^2(t)=0.1+0.01(1-e^{-2 t})^2$;
\item[{\bf  c.}]  $\lambda(t)=\lambda=0.4, \qquad\quad \sigma^2(t)=0.01(1.2+\sin t).$
\end{itemize}
We note that in all the cases the chosen functions are continuous and bounded. In particular, we choose $\lambda(t)$ constant or periodic to consider a seasonality effect of the infectious disease. Further,  in the {\bf Case~a} and {\bf Case~b},  $\lambda(t)$ periodically becomes negative, including situations in which a period of growth in the infection rate is followed by a period of regression of it. Three different choices are instead made for the function $\sigma^2(t)$: constant, asymptotically constant and periodic.\par
\begin{figure}[H]
\centering
\includegraphics[scale=0.4]{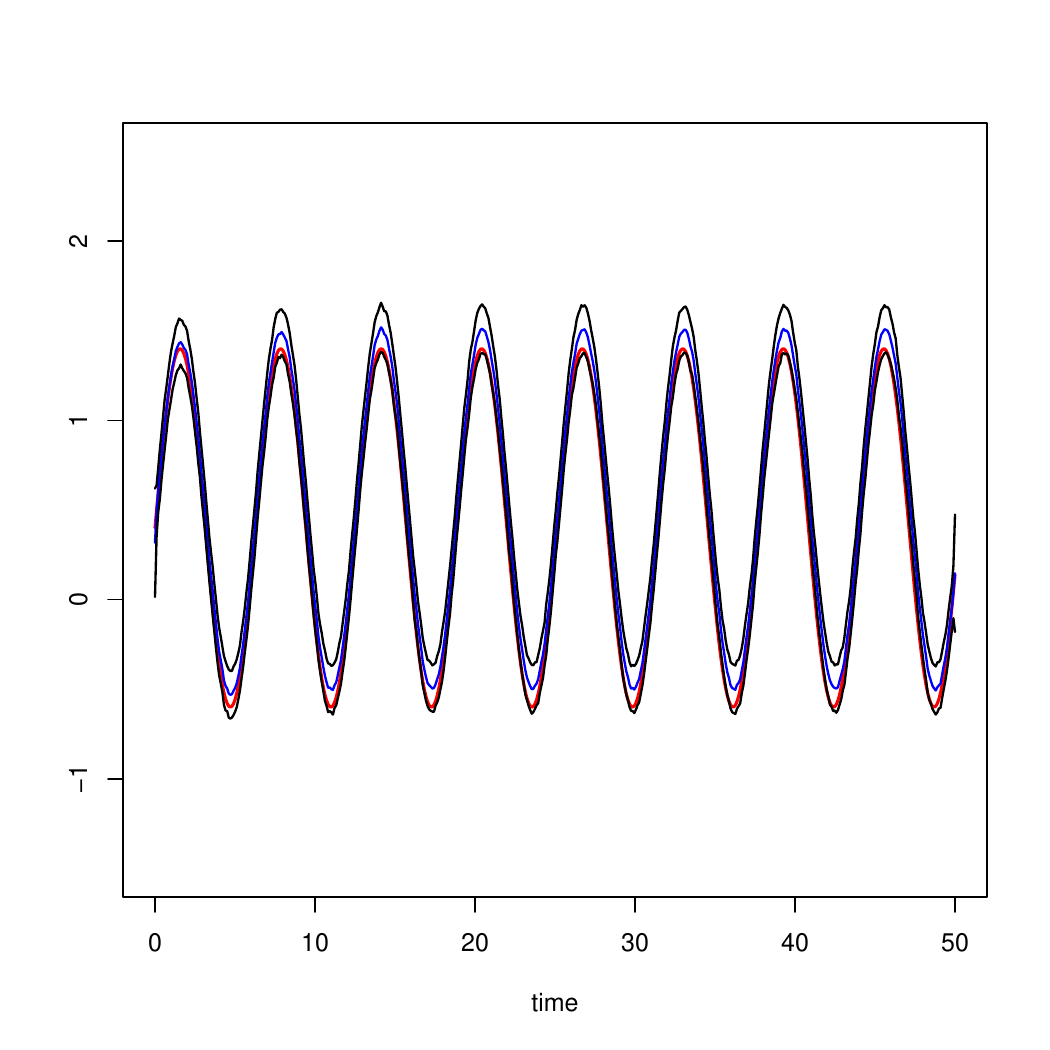}\includegraphics[scale=0.4]{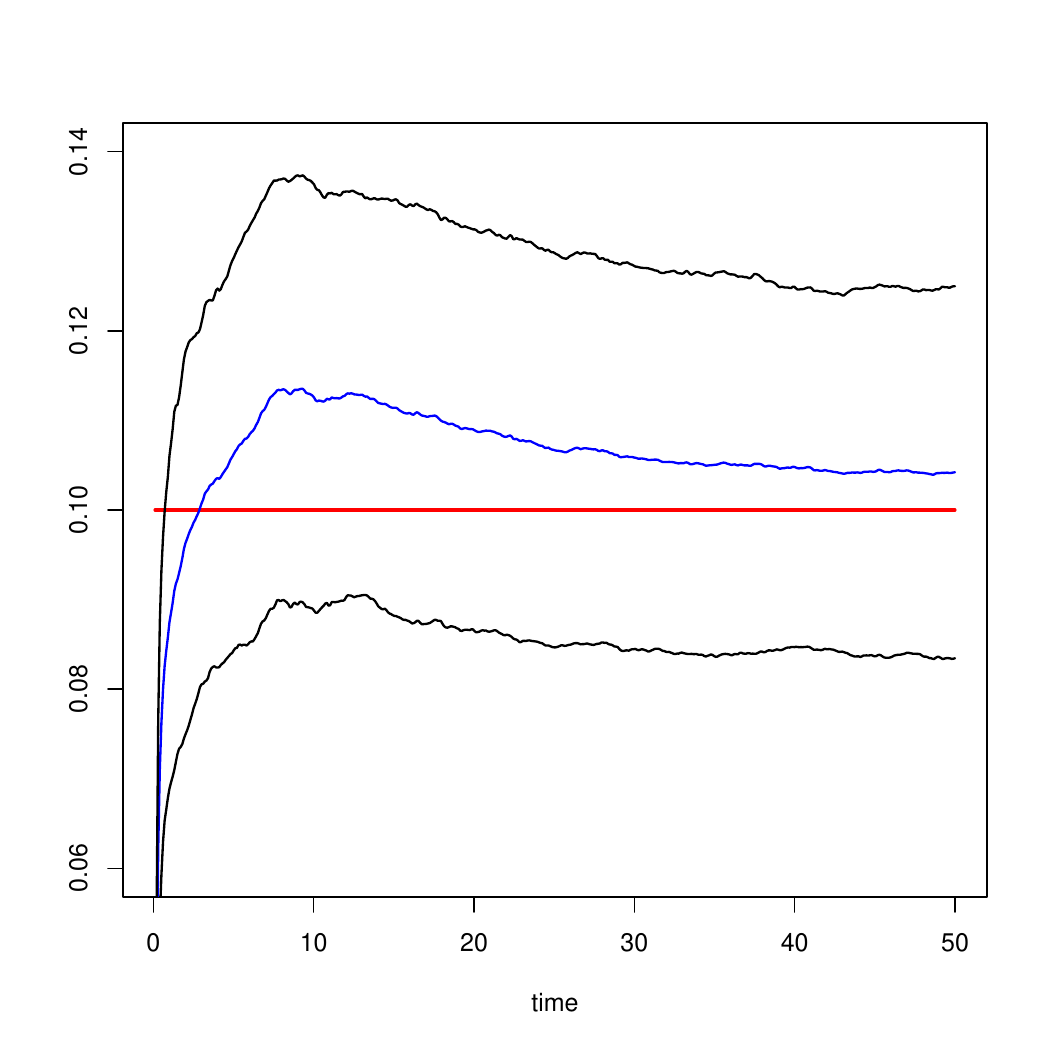}
\caption{Estimates of $\lambda(t)$ (on the left) and $\sigma^2(t)$ (on the rigth) for $\lambda(t)=0.4+\sin t$  and $\sigma^2(t)=\sigma^2=0.1$. The red curve is the true function, the red curve is the mean of the 500 obtained estimates. The black lines define  the observed confidence interval.}
\label{Fig2}
\end{figure}
\begin{figure}[H]
\centering
\includegraphics[scale=0.4]{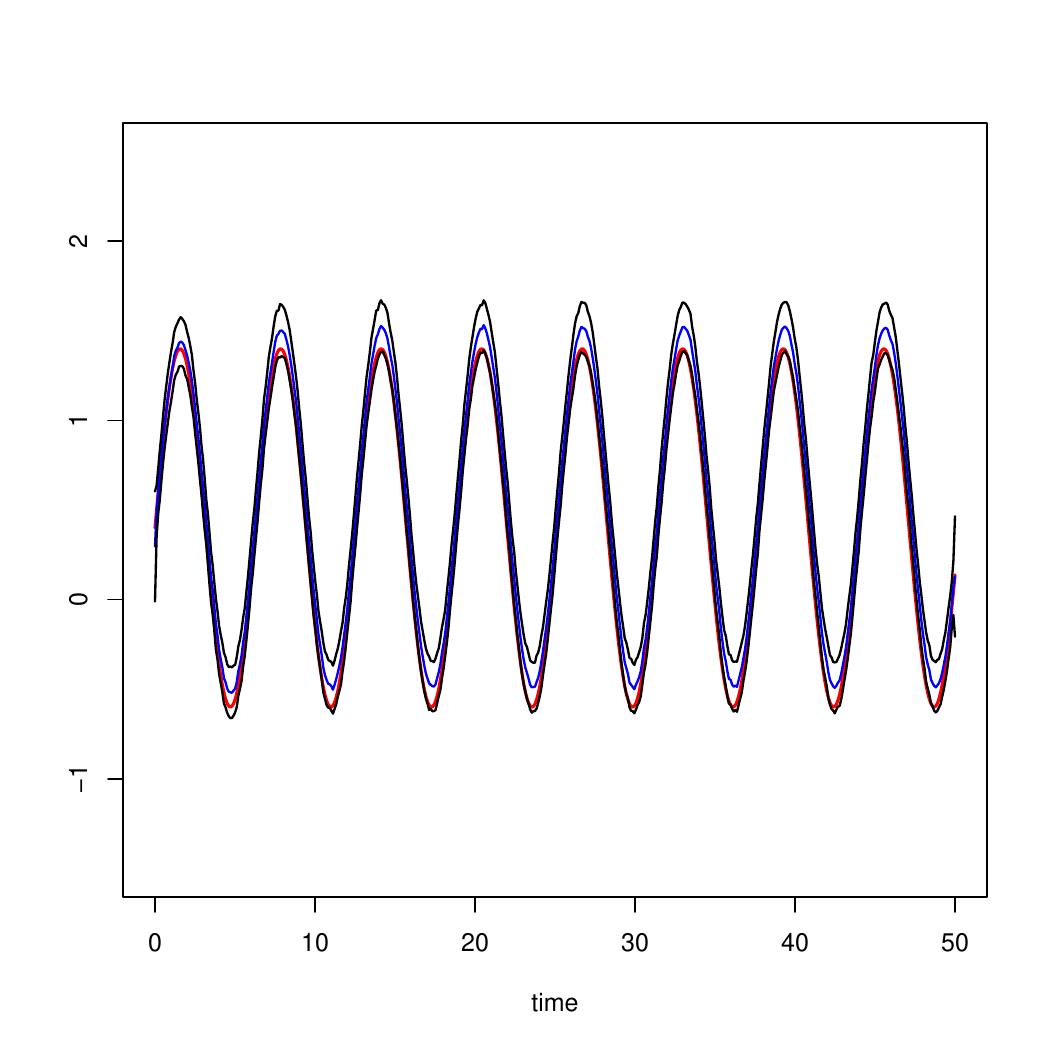}\includegraphics[scale=0.4]{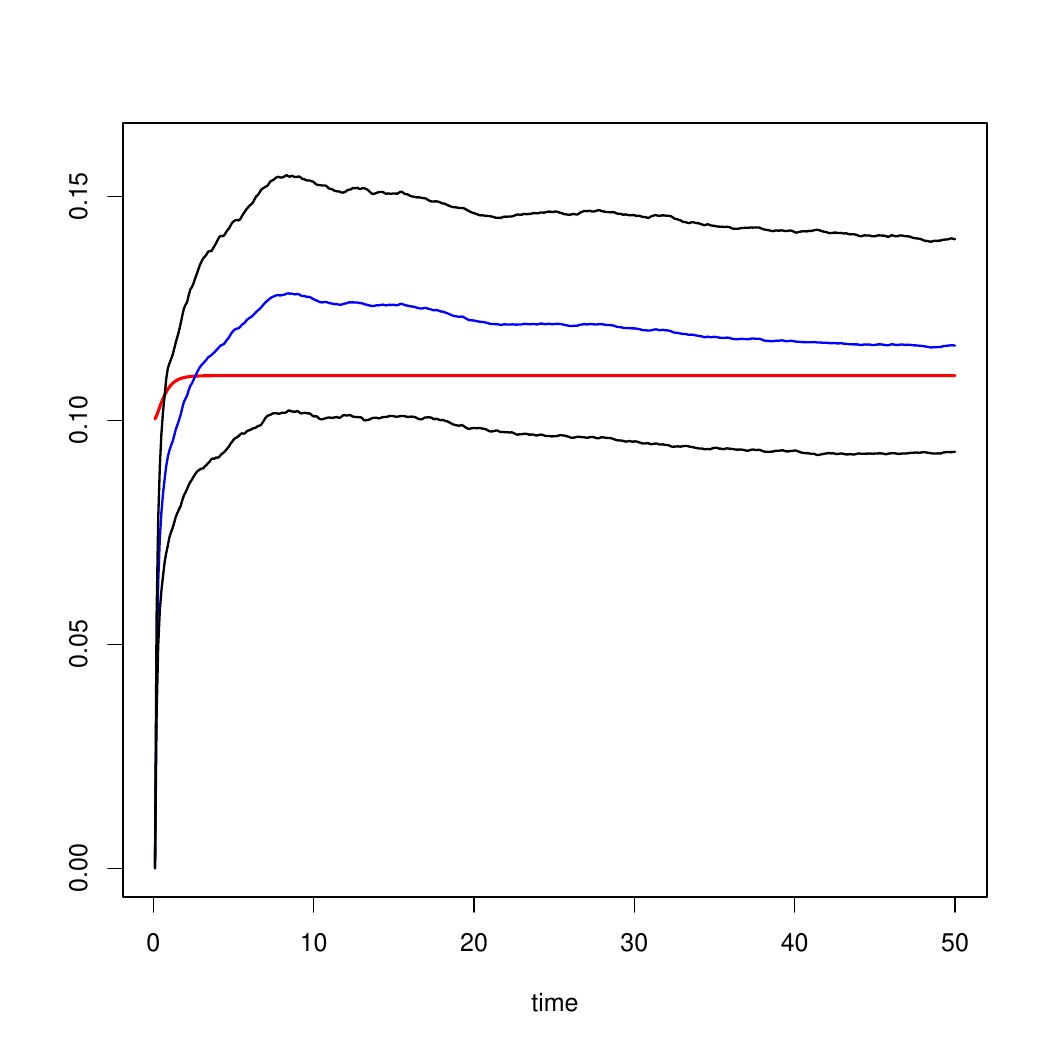}
\caption{As in Figure~\ref{Fig2} with $\lambda(t)=0.4+\sin t,$ and $\sigma^2(t)=0.1+0.01(1-e^{2 t})^2$. }
\label{Fig3}
\end{figure}
\begin{figure}[H]
\centering
\includegraphics[scale=0.4]{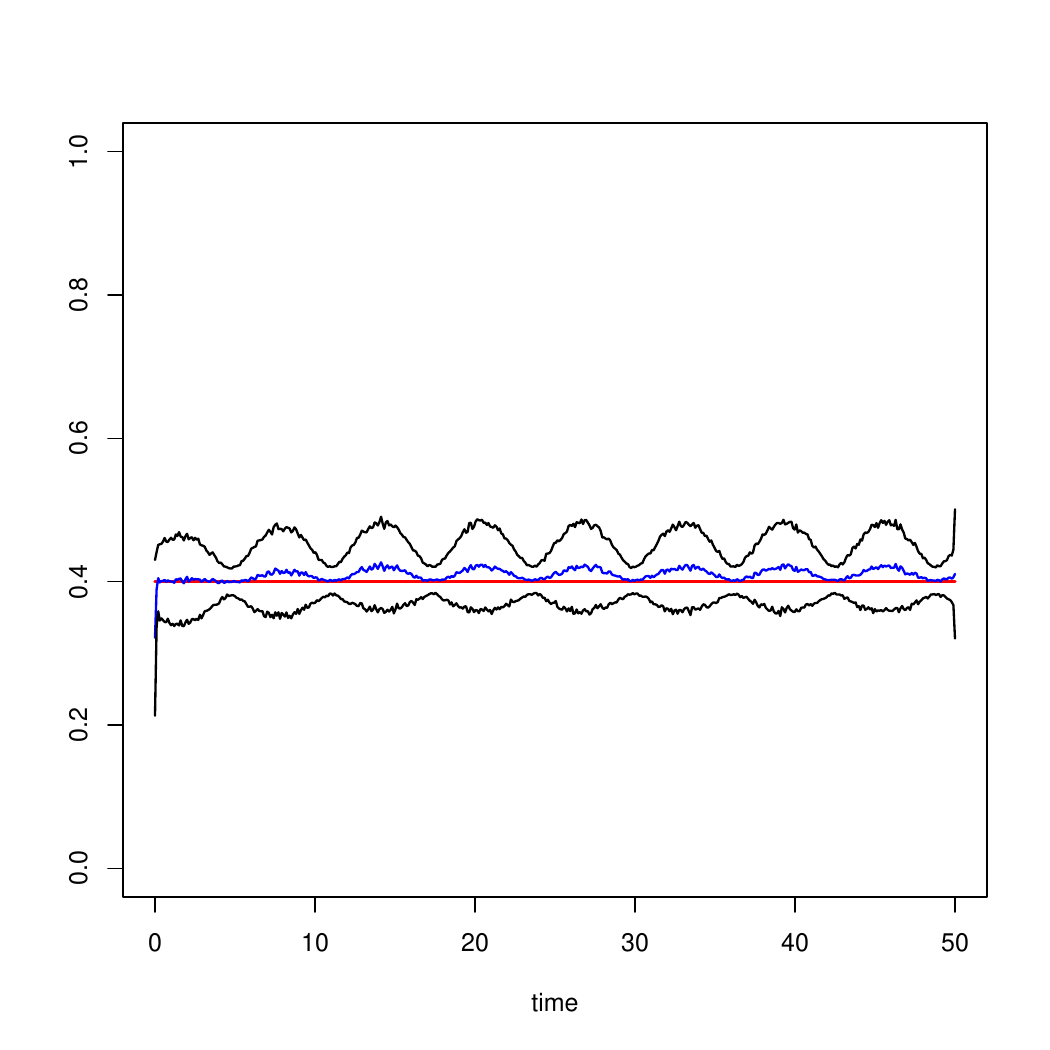}\includegraphics[scale=0.4]{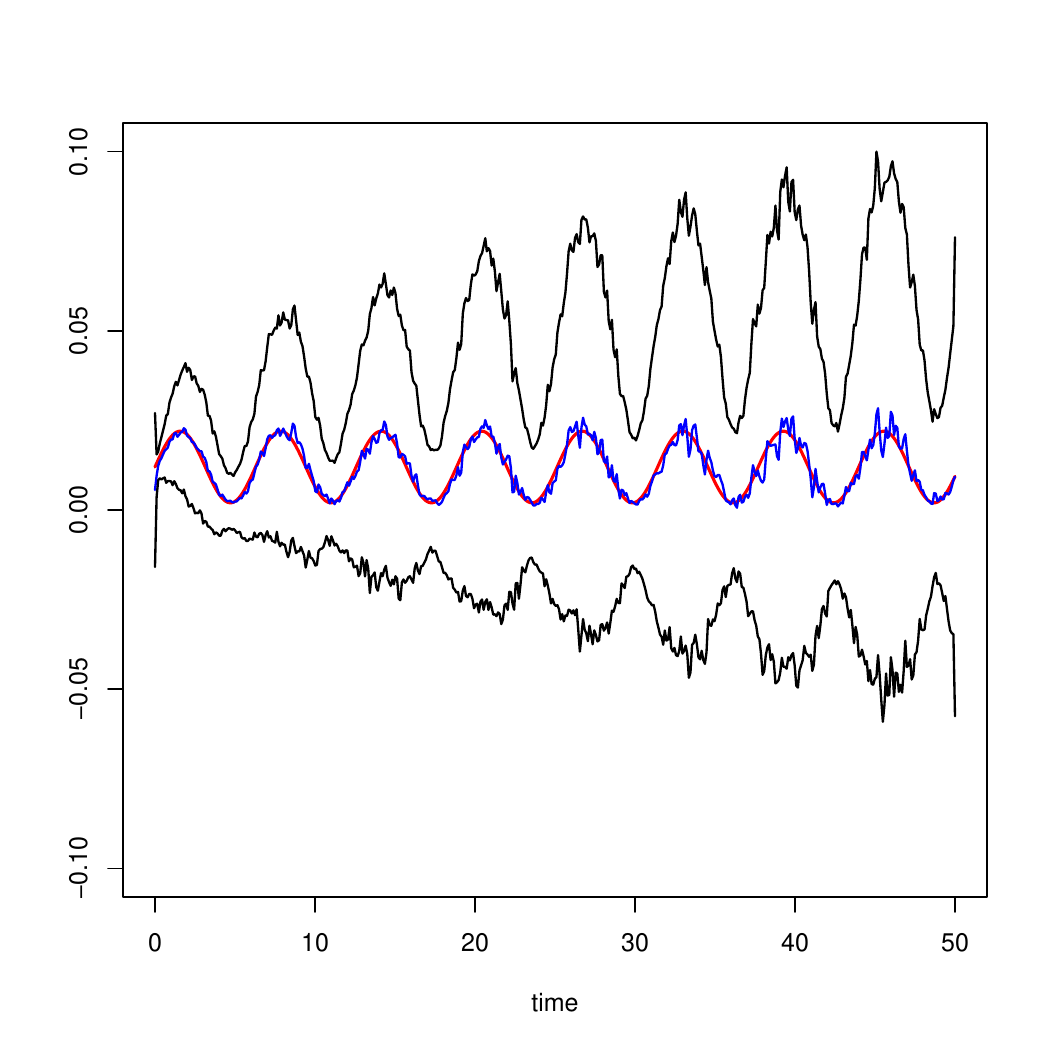}
\caption{As in Figure~\ref{Fig2} with $\lambda(t)=0.4$ and $\sigma^2(t)=0.01(1.2+\sin t)$. }
\label{Fig4}
\end{figure}

The results for the {\bf Case a} are shown in Figure~\ref{Fig2} for the functions $\lambda(t)$ (on the left) and for $\sigma^2(t)$ (on the right).
The red curve is the true function, the red curve is the mean of the 500 obtained estimates $\widehat\lambda_i(t)$ and $\widehat\sigma^2_i(t)$, i.e.  $\widehat\lambda(t)$ and $\widehat\sigma^2(t)$. The black lines represent the observed confidence interval for the functions $\lambda(t)$ and $\sigma^2(t)$ respectively, obtained as

$$
\widehat\lambda(t)\pm sd(\widehat\lambda(t)), \qquad \widehat\sigma^2(t)\pm sd(\widehat\sigma^2(t)),
$$
where
$$
sd(\widehat \lambda(t))=\sqrt{\frac{1}{500}\sum_{i=1}^{500}\big[\widehat\lambda_i(t)-\widehat \lambda(t)\big]^2}, \qquad sd(\widehat \sigma^2(t))=\sqrt{\frac{1}{500}\sum_{i=1}^{500}\big[\widehat\sigma_i^2(t)-\widehat\sigma^2(t)\big]^2}
$$
are the standard deviations of the estimators. We can observe on the left of Figure~\ref{Fig2}  that  $\widehat \lambda(t)$ fits very well the true function $\lambda(t)$ and also the amplitude of the confidence interval is small due to low values of the standard deviation. For the function $\sigma^2(t)$, we can see the the estimate $\widehat \sigma^2(t)$ decreases to the true value $0.1$ starting from the time $t=10$ and the confidence interval always contains the true value. \\
In the {\bf Case b} we consider the same periodic function of the {\bf Case a} for $\lambda(t)$, whereas $\sigma^2(t)$ is a time dependent function that asymptotically tends to the constant value $0.11$ and this value is quickly reached. This is the reason for which the results for the {\bf Case b}, illustrated in Figure~\ref{Fig3} for the function $\lambda(t)$ (on the left) and for $\sigma^2(t)$ (on the right), seem similar to the results in the previous case  (Figure~\ref{Fig2}),  although both the functions depend on time.\\
Very interesting are the results for the {\bf Case c} shown in Figure~\ref{Fig4}. Here the constant function $\lambda(t)=0.4$ is estimated, by using our procedure, by means a function $\widehat\lambda(t)$ showing a periodicity. Clearly, such periodicity is due to the presence of $\sin t$ in the function $\sigma^2(t)$. In this direction, we have to point out that the proposed  procedure for  estimating $\lambda(t)$ and $\sigma^2(t)$ is non parametric and the only assumption on the unknown functions is boundness of such functions. So, in the {\bf Case c}, the procedure, looking at the moments, and in particular to the mean and the covariance functions, captures the periodicity in the model and associates  it to both the functions $\lambda(t)$ and $\sigma^2(t)$. Anyway, the mean estimate $\widehat \lambda(t)$ is very close to the constant function $\lambda(t)=0.4$. Also the function $\widehat \sigma^2(t)$ fits very well the true function $\sigma^2(t)=0.01(1.2+\sin t)$ as shown in Figure~\ref{Fig4} on the right, although the confidence bands are increasingly further apart as time increases. These observations open the way to the possibility of finding a tool to discriminate between different estimated models. From this perspective, informational divergence could be a criterion to be used in a future study.

\section{An application to real data}
In this section, we apply our estimation procedure to the dataset {\bf twentymeas} included in the R package tsiR (see \cite{tsiRPackage}). It contains biweekly data (IP=2) related to measles infection for twenty locations in England from 1944-1964 and was studied in \cite{Rojal2010}. We point out that the application presented here is primarily for illustrative purposes of how the proposed procedure can be used with real data. People infected by measles become immune so more complicated models, such as SIR and SIRS, may be more adequate. However, even if simplified, we think that the SI model can also be used in this case. In fact, it is true that individuals who recover from measles become immune, but they remain infected, in a certain sense. The only difference is that they cannot infect other individuals. In our opinion, this aspect is taken into account with the fact that the transmission intensity function depends on time; in particular, we expect that $\lambda(t)$  becomes very small as time increases, showing that the disease is transmitted less frequently as recovered people have become immunized.

\begin{figure}[H]
\centering
\includegraphics[scale=0.4]{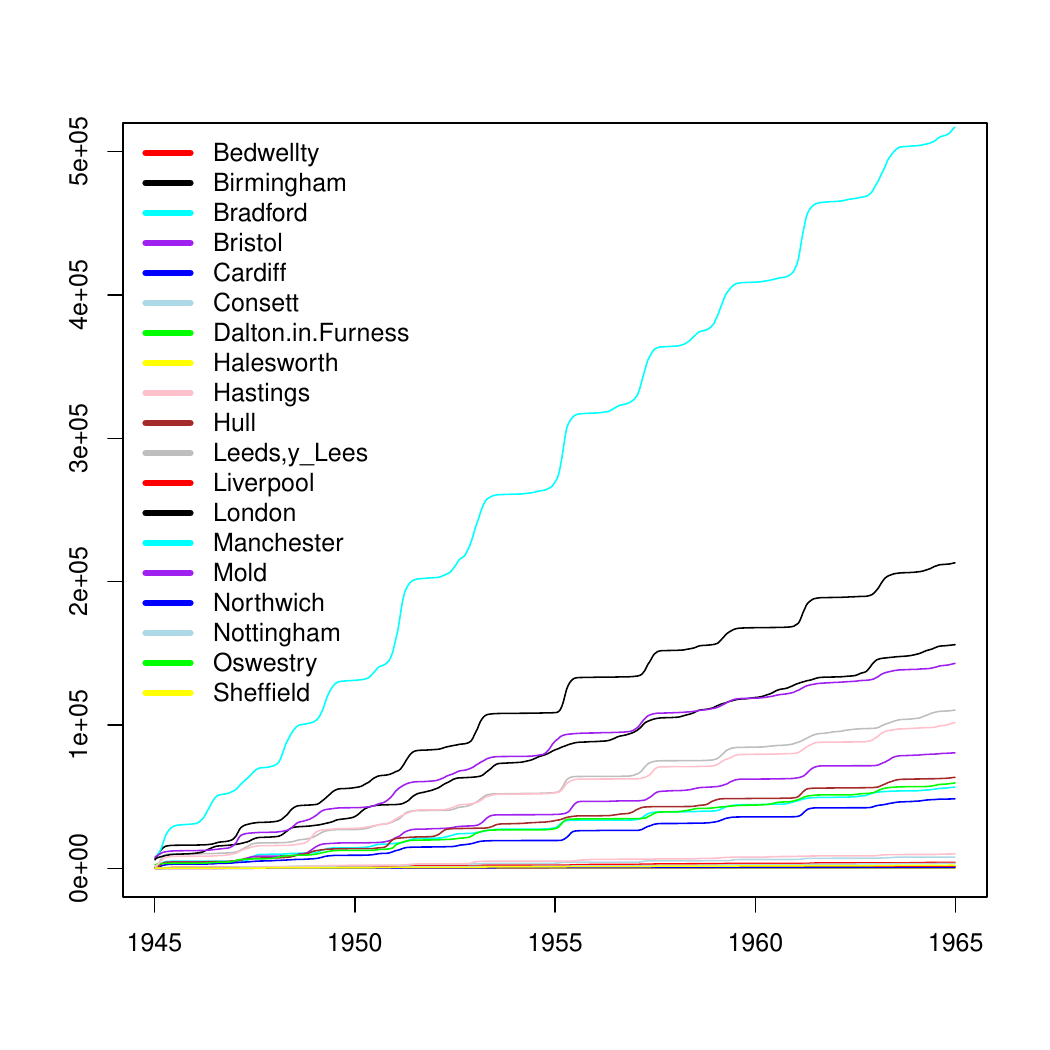}\includegraphics[scale=0.4]{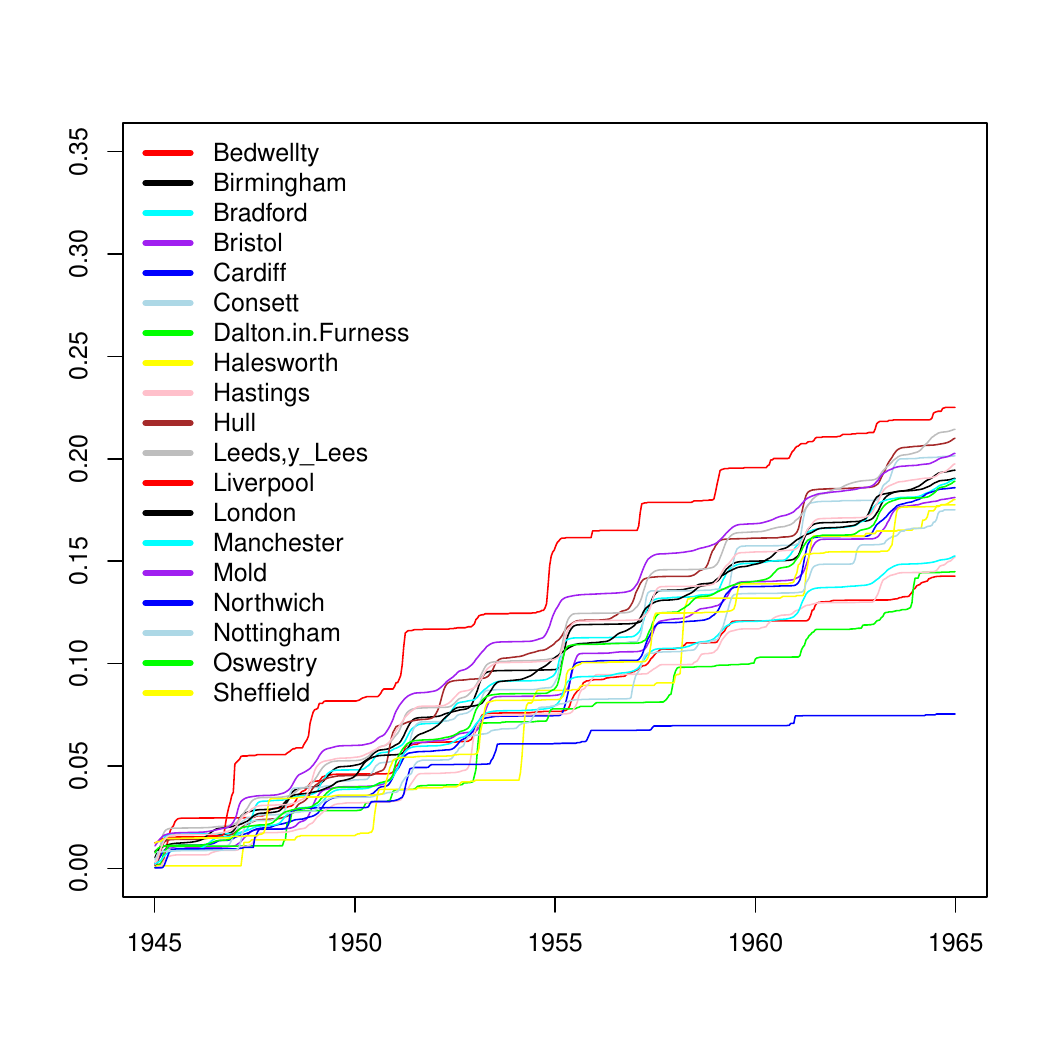}
\caption{Sample paths of the infected population (on the left) and  related  normalized sample paths (on the right).}
\label{Fig5}
\end{figure}

\begin{figure}[H]
\centering
\includegraphics[scale=0.4]{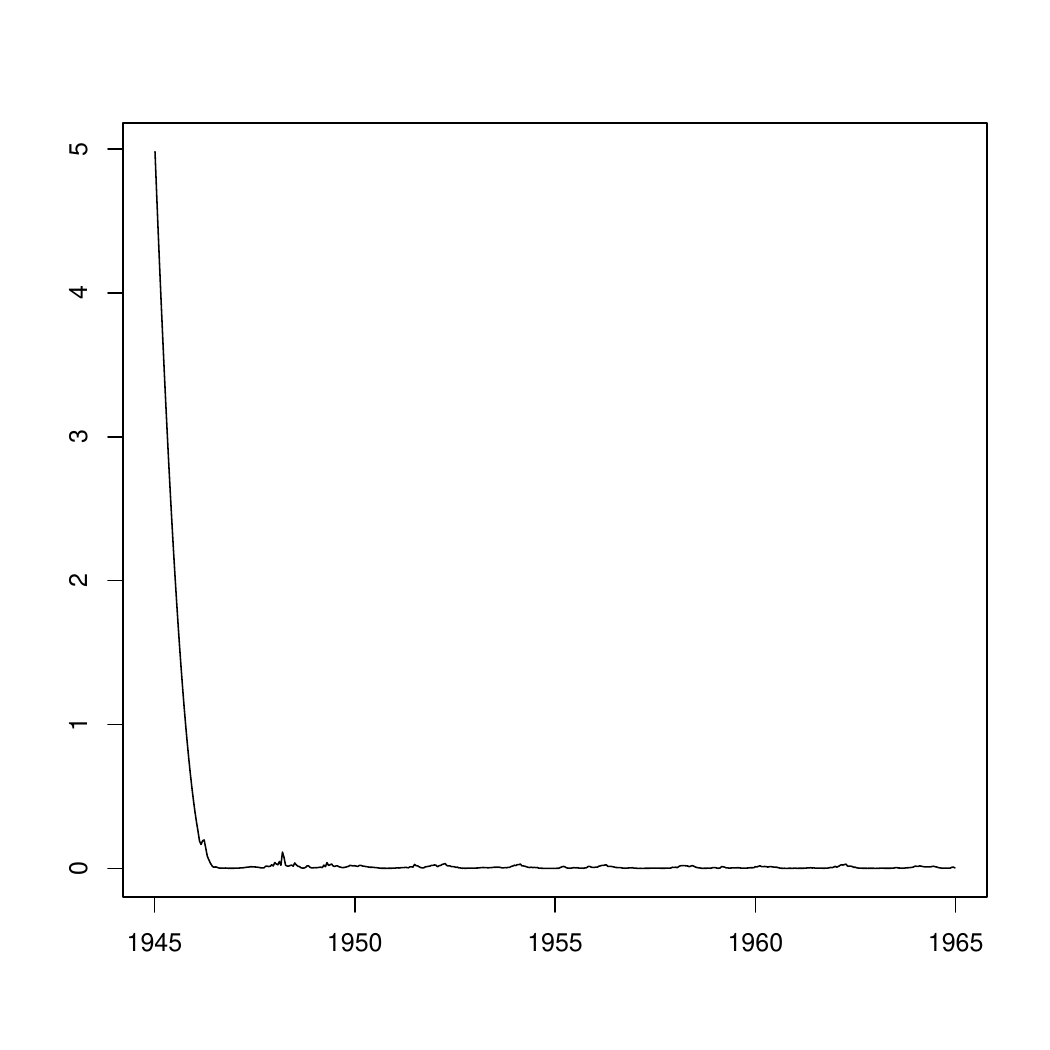}\includegraphics[scale=0.4]{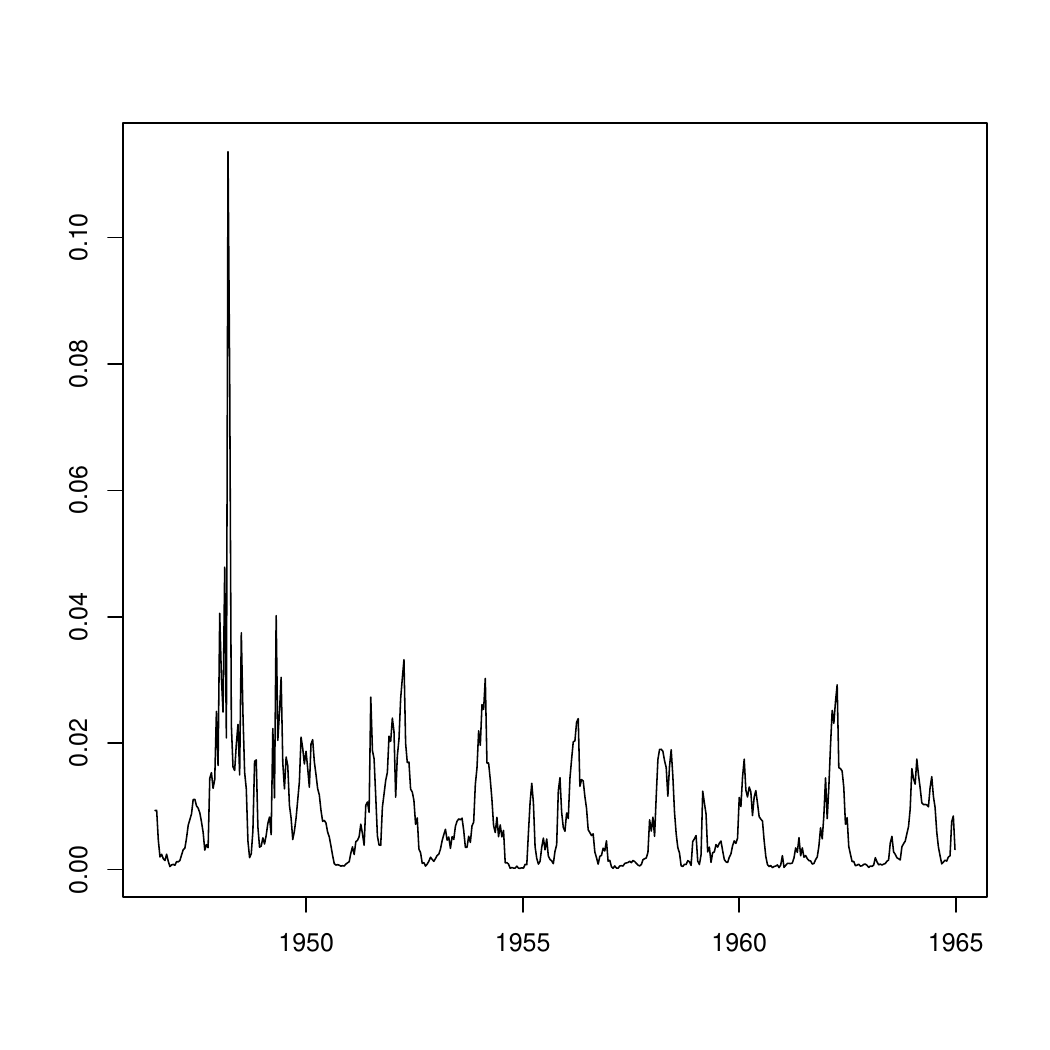}\\
\includegraphics[scale=0.4]{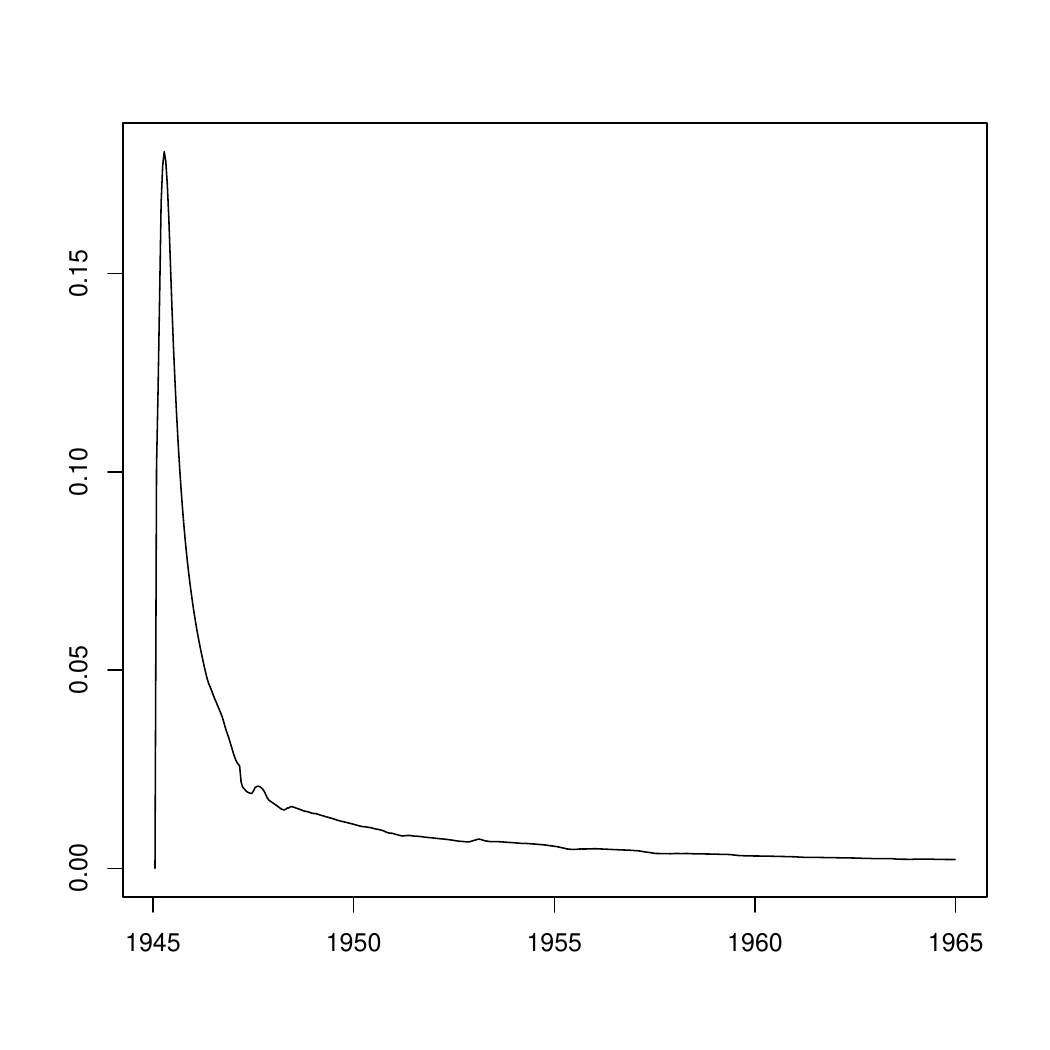}\includegraphics[scale=0.4]{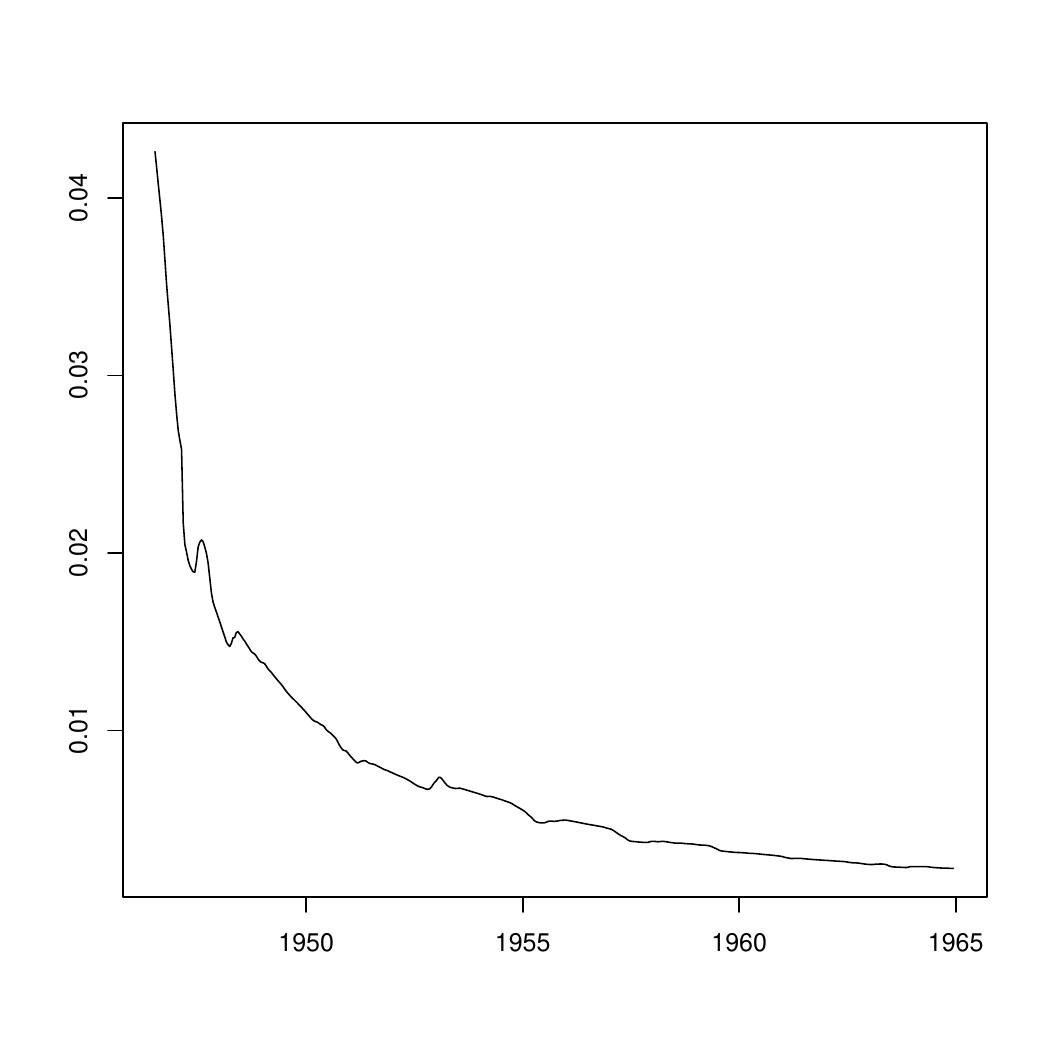}
\caption{Estimate of $\lambda(t)$ (on the top) and of $\sigma^2(t)$ (on the bottom) in the period 1944-1964 (on the left) and in the period 1946 to 1965 (on the right) for the normalized cumulative number of infected in England.}
\label{Fig6}
\end{figure}

For our analysis we firstly consider the cumulative number of infected for each location and we normalize this number by using the maximum number of population size, identified in the variable \lq\lq pop\rq\rq\ of the dataset. In Figure~\ref{Fig5} the sample paths of the infected population and the normalized sample paths are shown. Here we consider each normalized time series of the infected people as a sample path of a same diffusion process X(t) modeled via \eqref{stochasticsolution}. From the normalized process data (on the right of Figure~\ref{Fig5}), we fix $K=0.25$, i.e. the asymptotic infected population is $25\%$ of the total population. In Figure~\ref{Fig6} the estimated function $\widehat \lambda(t)$ (on the top) and $\widehat\sigma^2(t)$ (on the bottom) are plotted for the whole period of observation on the left, and for the period 1946 to 1965 on the right. By looking at $\widehat \lambda(t)$, we can see that $\widehat \lambda(t)$  presents a sharp decrease in the first period of observation due to its high initial followed by   a rapid tendency
to assume constant behavior over time. Further, in this  second period, the behavior of  $\widehat \lambda(t)$ shows a certain seasonality, as expected in infectious diseases. Also on the bottom an higher value of the estimate $\widehat\sigma^2(t)$ is observed in the initial period, after that it continues to decrease asymptotically to 0.

\section*{Conclusion}
We have considered a stochastic diffusion process for the time-inhomogeneous deterministic SI epidemic model and we have proposed an estimating procedure to inferring the considered process. A relevant issue concerning the model under consideration is the fact that it works quite well in situations where the total population size is large. In situations where the size is sufficiently small, several authors recommend the use of discrete state space processes. However, it should be noted that the proposed estimation procedure allows dealing when the population size is not too large. An example of this can be found in the example described at the beginning of Section 4, where the total size does not differ so much from the original (200 vs. 20).

Another interesting aspect is related to the assumption that the total population size is constant, which is usually a generalized assumption in this type of models.This is due to the fact that they reflect a study carried out over a period of time that is not too long to consider relevant variations in the size of the population. One line of future work is to consider a population growth pattern, either Logistic, Gompertz or other growth models.

To estimate the functions $\lambda(t)$ and $\sigma^2(t)$ that characterize the process, we have proposed a procedure based on the GMM method  combined with the interpolation of the sample mean and covariance points, so
the consistence of the  estimators derives from the consistence of the GMM estimator and   the uniform convergence of the used interpolation method.
It should be noted that the proposed procedure does not make any assumptions about the functional form of the unknown functions. The only necessary assumption is that the functions involved are continuous and bounded in the observation interval. Several simulation studies have been carried out which demonstrate the validity of the proposed methodology.

The results for both the estimates obtained in the considered simulation studies seem very closed to the \lq\lq true\rq\rq\  functions. In particular, for the time homogeneous case, we have compared the results obtained via the MLE procedure with those ones obtained with our procedure: concerning the parameter $\lambda$, representative of the infection rate, the estimates are very closed with those obtained via our method,
some  differences are instead found for the parameter $\sigma^2$, related to environmental variability, in which better performances are shown for the MLE due to the strong assumption of the constancy of the parameters. However, the MRE's for the two methods and for the parameters are in all the cases comparable. Other simulation cases have been  analyzed in which the functions $\lambda(t)$ and/or $\sigma^2(t)$ are time-dependent with particular reference to situations in which seasonality effects  of  the dynamics of the infection are included. Finally, we have applied our estimation procedure to the dataset {\bf twentymeas} included in the R package tsiR (see \cite{tsiRPackage}). It contains biweekly data (IP=2) related to measles infection for twenty locations in England from 1944-1964. Regarding the estimates obtained for $\lambda(t)$ and $\sigma^2(t)$ via our method, we observe a sharp decrease in the first period of observation due to their high initial values, followed by a rapid tendency to assume constant behavior over time. Further, in this second period, the behavior of the transmission intensity function shows a certain seasonality, as expected in infectious diseases. Instead,  the estimate of $\widehat \sigma^2(t)$  rapidly decreases to a constant value without showing any periodicity.

%
%
\section*{Conflict of interest}
Giuseppina Albano is an special issue editor for Mathematical Biosciences and Engineering and
was not involved in the editorial review or the decision to publish this article. All authors declare that
there are no competing interests.

\section*{Use of AI tools declaration}
The authors declare they have not used Artificial Intelligence (AI) tools in the creation of this article.

\section*{Acknowledgments} 
This research is partially supported by MUR-PRIN 2022, project 2022XZSAFN \lq\lq Anomalous Phenomena on Regular and Irregular Domains: Approximating
Complexity for the Applied Sciences\rq\rq (Italy), by MUR-PRIN 2022 PNRR, project P2022XSF5H \lq\lq Stochastic Models in Biomathematics and Applications\rq\rq  (Italy), by PID2020-1187879GB-100 and CEX2020-001105-M grants, funded by MCIN/AEI/10.13039/501100011033 (Spain). G. Albano and V. Giorno are members of the
GNCS-INdAM.

\end{document}